\documentclass[%
reprint, twocoloumn,
amsmath,amssymb,
aps,
]{revtex4-2}
\usepackage{graphicx}% Include figure files
\usepackage{dcolumn}% Align table columns on decimal point
\usepackage{blindtext}
\usepackage{bm}
\usepackage{tikz}
\usepackage{pgfplots}
\pgfplotsset{compat=1.14}
\usepackage{adjustbox}
\usetikzlibrary{decorations.pathreplacing}
\usepackage{amssymb,float}
\usepackage{float}
\usepackage{amsmath}
\usepackage{lipsum}
\usepackage{rotating}
\usepackage{color}
\usepackage{xcolor}
\usepackage{pgfplots}
\usepackage{soul}
\usepackage{cancel}

\DeclareRobustCommand{\orcidicon}{%
	\begin{tikzpicture}
	\draw[lime, fill=lime] (0,0) 
	circle [radius=0.16] 
	node[white] {{\fontfamily{qag}\selectfont \tiny ID}};
	\draw[white, fill=white] (-0.0625,0.095) 
	circle [radius=0.007];
	\end{tikzpicture}
	\hspace{-2mm}
}

\foreach \x in {A, ..., Z}{\expandafter\xdef\csname orcid\x\endcsname{\noexpand\href{https://orcid.org/\csname orcidauthor\x\endcsname}
		{\noexpand\orcidicon}}
}

\DeclareRobustCommand{\orcidicon}{%
	\begin{tikzpicture}
	\draw[lime, fill=lime] (0,0) 
	circle [radius=0.16] 
	node[white] {{\fontfamily{qag}\selectfont \tiny ID}};
	\draw[white, fill=white] (-0.0625,0.095) 
	circle [radius=0.007];
	\end{tikzpicture}
	\hspace{-2mm}
}

\foreach \x in {A, ..., Z}{\expandafter\xdef\csname orcid\x\endcsname{\noexpand\href{https://orcid.org/\csname orcidauthor\x\endcsname}
		{\noexpand\orcidicon}}
}

\usepackage{etoolbox}
\apptocmd{\sloppy}{\hbadness 10000\relax}{}{}%solve the warn"Underfull \hbox (badness 3323)"

\usepackage{silence}
\newcommand{\Nel}{N_{\text{el}}}

\newcommand{\la}{\langle}
\newcommand{\ra}{\rangle}
\newcommand{\rt}{\right}
\newcommand{\lf}{\left}

\newcommand{\kp}{{\mathbf k}}

\renewcommand{\i}{\text{in}}

\newcommand{\br}{\mathbf r}

\newcommand{\eq}[1]{Eq.~(\ref{#1})}

\newcommand{\Hint}{\hat H_{\text{int}}}

\usepackage{hyperref}
\usetikzlibrary{animations}
\begin{document}
	%	\title{Theoretical description of x-ray absorption spectroscopy of excitons with Bethe-Salpeter equation}
	\title{Revealing fingerprints of valence excitons in x-ray absorption spectra with the Bethe-Salpeter equation}
	%\author{Nasrin Farahani${}^{1}$}
	\author{Nasrin Farahani${}^{1}$ \orcidA{}}
	\author{Daria Popova-Gorelova${}^{1,2,3}$\orcidB{}}
	\email{darya.gorelova@uni-hamburg.de}
	\affiliation{${}^{1}$I. Institute for Theoretical Physics and Centre for Free-Electron Laser Science,
		Universit\"at Hamburg, Notkestr.~9, 22607 Hamburg, Germany\\
		${}^{2}$The Hamburg Centre for Ultrafast Imaging (CUI),Luruper Chaussee 149, 22607 Hamburg, Germany\\
                 ${}^3$Institute of Physics, Brandenburg University of Technology Cottbus-Senftenberg, Erich-Weinert-Stra\ss e 1, 03046 Cottbus, Germany
	}
	
	\begin{abstract}

		{The Bethe-Salpeter equation (BSE) is a powerful theoretical approach that is capable to accurately treat electron-hole interactions in materials in an excited state. We developed an ab initio framework based on the BSE to describe a pump-probe experiment, in which an x-ray pulse probes solid-state valence excitons by means of x-ray absorption spectroscopy. Our theoretical framework is of relevance for an accurate modeling of pump-probe experiments of photo-excited materials that utilize novel capabilities offered by x-ray science.}

	\end{abstract}
	
	\maketitle
	
	%\blindtext \cite{article-minimal}
	
	%\bibliographystyle{apsrev4-1} % Tell bibtex which bibliography style to use
	
	%\bibliography{xampl} % Tell bibtex which .bib file to use (this one is some example file in TexLive's file tree)

	%	{
	%		\section {INTRODUCTION}}
	%{\color{red} Developing an ab initio description of X-ray absorption spectroscopy to probe valence-excitons in solids.}\\
	\section{INTRODUCTION}
	Excitons, bound electron-hole pairs, can be created in a material due to absorption of light and are responsible for energy-conversion processes. Excitonic effects can play an essential role for electronic excitations in solids and influence their dynamics. Despite the crucial role of excitons for functionality of optoelectronic devices, their properties are not completely understood. X-ray absorption spectroscopy (XAS) is an element-specific technique, in which an x-ray pulse resonantly excites an electron from a core orbital into unoccupied states of a material or a molecule \cite{de2008core, yano2009x, baumgarten2012x}.  Its ability to deliver information about electronic structure of matter made it a powerful method for studying materials in an excited state \cite{KrupaARPC16, santomauro2017localized, fracchia2018time, HuangInno21}. Especially, it has been applied to investigate dynamics of excitons in functional materials \cite{garratt2021ultrafast, garratt2022direct, hillyard2009atomic,  palmieri2020mahan, ZuerchNatComm17, GarrattStructuralDynamics24}. Accurate interpretation of features in XAS spectra and mapping them to properties of a photo-excited material relies on an accurate ab initio modelling of corresponding pump- and probe-induced processes. 
	
       In this work, we present an ab initio framework based on the Bethe-Salpeter equation (BSE) \cite{salpeter1951relativistic, strinati1988application, onida2002electronic} to describe an experiment in which a pump pulse creates a valence exciton in a material and a probe x-ray pulse probes the exciton by inducing a transition of a core electron into unoccupied states below the Fermi level. The BSE is a state-of-the-art ab initio many-body approach that can accurately treat excited-states in matter including excitonic effects. On the one hand, it has been proven to be a powerful tool for calculations of optical excitations in solids and for the description of valence excitons. On the other hand, this method substantially improves calculations of XAS in solid-state materials since excitonic effects can also play an important role for core-excited states. It is common to use the BSE for calculations of either optical absorption spectra, when an initial state is a ground state and final states are optically-excited states  \cite{hirose2015all, gui2018accuracy, bruneval2015systematic, jacquemin2017bethe}, or x-ray absorption spectra, when an initial state is a ground state and final states are core-excited states \cite{gulans2014exciting, draxl2017exciting}. The BSE was employed to describe an optical-pump--optical-probe experiment that involves valence-excited states \cite{sangalli2023exciton}. In our study, we employ the BSE to describe initial states that are optically-excited states and to describe final states that are core-excited states.
	
	Several other theoretical approaches were recently applied to describe XAS of excitons in materials. This includes a formalism based on the Bloch model \cite{picon2019attosecond, CistaroJChThComp23}, formalism based on the Time-Dependent Density Functional Theory \cite{garratt2022direct, ZuerchNatComm17} and also treatment using model Hamiltonians \cite{malakhov2023exciton, hansen2023theoretical}. The Bethe-Salpeter equation was employed in a study of extreme ultraviolet spectra of an optically-excited material, where core-excited states were obtained by the BSE and valence-excited states were treated by adjusting carrier occupation \cite{vinson2022advances,KleinJPC23, KleinJACS22}. Approaches that are applicable to molecular systems are also being developed \cite{DutoiPRA13, SkeidsvollPRA20, KhaliliStructuralDynamics20, RottStructuralDynamics21}. The advantage of our approach is that it accurately treats many-body effects for both valence and core excitations in materials taking the effect of electron-hole couplings into account.

	We illustrate our results with calculations for 4H-SiC that is known to have a strong excitonic effect \cite{bockstedte2010many,klahold2018high, klahold2020band}. One advantage of this material is that its optical absorption spectra have been extensively investigated \cite{zhang2023phonon, demmouche2018electronic, peng2004theoretical, de2003spectroscopy, adolph1997optical, ahuja2002optical, ching2006electronic, gao2015band, lindquist2001ordinary, harris1995properties} and its XAS has been measured \cite{tallarida2006x, luning1999electronic}, which allows us to verify the accuracy of the BSE calculations. The other advantage is that it has two types of atoms in the unit cell and we can compare results for Si and C K-edges.

	The article is organized as follows. Sec.~\ref{SecMethod} presents our derivations and provides computational details. We illustrate our methodology using calculations for 4H-SiC and discuss the results in Sec.~\ref{SectionResults}. Sec.~\ref{SectionConclusions} concludes the study.
	
	\newpage
	\section{Methodology}
	\label{SecMethod}

	{ An x-ray absorption threshold (XAT) is the energy, below which an x-ray absorption signal from a system with all states occupied below the Fermi level is vanishing. The energy region below XAT is also referred to as a pre-edge region of XAS. In our article, we consider a two-step process, in which an optical excitation creates vacancies below the Fermi level. As a result, an interaction with an x-ray pulse with an energy below XAT leads to a signal that emerges in the pre-edge region. Such an additional peak in the pre-edge region of XAS after an optical excitation has been observed in experimental studies  \cite{garratt2021ultrafast, garratt2022direct, hillyard2009atomic, ZuerchNatComm17, ParkNatComm22, MilneStucturalDynamic23}. In our article, we derive the framework to calculate this signal.
	
	In the two-step process considered, an initial state is a ground state, an intermediate state is a valence-excited state and a final state is a core-excited state as shown in Fig.~\ref{Fig_Pump-Probe-Scheme}. We treat both valence- and core-excited states as two-particle wave functions that can be obtained in a BSE calculation. A standard implementation of a BSE involves calculation of a static microscopic dielectric matrix in the random phase approximation for a calculation of a screening. In the considered process, a transition to a core-excited state is from a valence-excited state, which is not static. This still does not lead to an additional approximation, because an obtained core-excited state is an eigenstate of a many-body electron system independently of an intermediate state.
	
	 %Due to the action of an optical pulse, states above Fermi level becomes occupied.
	 Another possible final state after the two-step process is a doubly-excited state with a valence and a core excitation. Such a final state can be reached, if the energy of an x-ray pulse is above XAT. This signal would appear in the spectral region, where stationary XAS signal is nonzero. We cannot describe a doubly-excited state taking excitonic effects into account and that is why we cannot address changes of an x-ray signal above XAT in our formalism. In our study, we focus on the signal in the energy region below the XAT.
	
	\subsection{Analysis}

	We derive XAS of a photo-excited material within the second-order time-dependent perturbation theory and second quantization formalism \cite{santra2008concepts} in Appendix A. { Our approach is valid, if the interaction with each of pulses, an optical and an x-ray one, is in a linear regime. For the derivation of the cross section, we assumed that pulses do not overlap in time. We also assume that the time delay between the pulses is much shorter than a decoherence time due to coupling with phonons, which is on a time scale of hundreds of femtoseconds for solid-state systems. The finite life times of optically-excited and core-excited states are taken into account by the broadening parameters $\Gamma_I$ and $\Gamma_F$.} The absorption cross section in the pump-probe experiment is then proportional to
	\begin{widetext}
		\begin{align}
			\sigma\propto\sum_{\lambda_F}\lf |  \sum_{\lambda_I}\frac{ \la \Psi_{\lambda_F}  |\hat \psi^\dagger(\br_1) e^{i\mathbf k_x\cdot \br_1}(\boldsymbol\epsilon_x\cdot \hat{\mathbf p})\hat \psi (\br_1) |\Psi_{\lambda_I}\ra }{\Gamma_F - i(E_{\lambda_I} - E_{\lambda_F}+\omega_x)} 
			\frac{  \la  \Psi_{\lambda_I}|  \hat \psi^\dagger(\br_2) (\boldsymbol\epsilon_{\text{o}}\cdot \hat{\mathbf p})\hat \psi (\br_2) | \Psi_0\ra}{\Gamma_I - i(E_{0} - E_{\lambda_I}+\omega_{\text{o}})}  \rt|^2, \label{Eq_XAS_general}
		\end{align}
	\end{widetext}
	where this and all the following expressions are in atomic units. $|\Psi_0\ra$ is the ground state of a many-electron system with an energy $E_0$, $|\Psi_{\lambda_I}\ra$ is its optically-excited state with an energy $E_{\lambda_I}$,  $|\Psi_{\lambda_F}\ra$ is its core-excited state with an eigenenergy $E_{\lambda_F}$. $\mathbf k_x$ is the wave vector of the x-ray pulse. $\omega_x$, $\boldsymbol\epsilon_x$, and $\omega_{\text{o}}$, $\boldsymbol\epsilon_{\text{o}}$ are the photon energies and polarizations of an x-ray probe pulse and an optical pump pulse, respectively. $c$ is the speed of light, $\hat{\mathbf p}$ is the momentum operator and $\psi^\dagger$, $\psi$ are the field operators \cite{fetter2012quantum}. Within the BSE formalism, many-body wave functions are eigenfunctions of a two-particle excitonic Hamiltonian that are represented as  \cite{quintela2022theoretical, karni2021moir, man2021experimental} 
	\begin{align}
		|\Psi_\lambda \ra = \sum_{e,h,\kp} A^\lambda_{e h\kp}| e_\kp h_\kp \ra,\label{eq_wave_exciton}%=  \sum_{e_\kp,h_\kp} A^\lambda_{vc}c_{e_\kp}^\dagger \hat c_{h_\kp} | \Psi_0\ra,  
	\end{align}
	where $ | e_\kp h_\kp \ra$ is a basis state, where an electron occupies a one-particle state $|\phi_{e\kp}\ra$ and a hole occupies a one-particle state $|\phi_{h\kp}\ra$. We can also represent this state as $ | e_\kp h_\kp \ra = \hat c_{e\kp}^\dagger \hat c_{h\kp} | \Psi_0\ra$, where  $ \hat c_{e\kp}^\dagger$ ($ \hat c_{h\kp}$) creates (annihilates) an electron from the one-particle state $|\phi_{e\kp}\ra$ ($|\phi_{h\kp}\ra$) of the many-body ground-state wave function $| \Psi_0\ra$. If many-body wave functions can be represented as a product state of the Fock space and one-particle wave functions form an orthonormal basis set, then the field operators can be represented as $\hat \psi (\mathbf r)=\sum_{\mu,\kp} \hat c_{\mu\kp}\phi_{\mu\kp}(\br)$ and $\hat \psi^\dagger (\mathbf r)=\sum_{\mu,\kp} \hat c^\dagger_{\mu\kp}\phi^\dagger_{\mu\kp}(\br)$ \cite{cistaro2022theoretical}. Within a standard application of the BSE  for solid-state materials, $\phi_{\mu\kp}(\br)$ are Kohn-Sham orbitals, where $\mu$ is a band and a spin index, and $\kp$ is a Bloch vector. We now assume that the wave functions of both the initial state $\lambda_I$ and the final state $\lambda_F$ that enter \eq{Eq_XAS_general} are the solutions of the BSE equation and derive the following expression for the absorption cross section in Appendix A
	% In general, one-particle wave functions can be selected to be Hartree-Fock orbitals as well \cite{blase2020bethe}.  
	\begin{widetext}
		\begin{align}
			\sigma = 
			\frac1c\sum_{\lambda_F} 
			\lf| \sum_{\lambda_I} w_{\lambda_I} \frac{\sum_{e,v,\mu,\kp } A^{*\lambda_F}_{e \mu\kp} A^{\lambda_I}_{e v\kp} \int d^3 r \phi_{v\kp}^\dagger(\br)e^{i \mathbf k_x\cdot\br}( \boldsymbol\epsilon_x \cdot \hat {\mathbf p})  \phi_{\mu\kp}(\br)}{\Gamma_F - i(E_{\lambda_I} - E_{\lambda_F}+\omega_x)}
			\rt|^2.\label{Eq_XAS_BSE}
		\end{align}
	\end{widetext}
	Here,  $\mu$ are core states and  $v$ are valence states. $w_{\lambda_I}$ is a ``weight" associated with a valence-excited state $\lambda_I$ 
	\begin{align}
		w_{\lambda_I} & = \sqrt{\frac{2\pi I_{\text{o}}}{c\omega_{\text{o}}^2}}\frac{ \sum_{e,h,\kp} A^{*\lambda_I}_{eh\kp}\int d^3 r \phi_{e\kp}^\dagger(\br)( \boldsymbol\epsilon_{\text{o}} \cdot \hat {\mathbf p})  \phi_{h\kp}(\br) }{\Gamma_I - i(E_{0} - E_{\lambda_I}+\omega_{\text{o}})},
	\end{align}
	where $I_{\text{o}}$ is the intensity of the pump pulse. The expression in \eq{Eq_XAS_BSE} can be intuitively understood as the cross section resulting due to x-ray-induced transitions from states $\lambda_I$ to states $\lambda_F$ weighted by probability amplitudes of an optical excitation into the states $\lambda_I$.

The expression in \eq{Eq_XAS_BSE} has similarities to the description of resonant inelastic x-ray scattering (RIXS) using the BSE described in Refs.~\cite{vorwerk2020excitation, Shirley.RIXS, Vinson.RIXS}.} { RIXS is also a two-step process, which involves absorption and emission of an x-ray photon. An initial state in the process considered in these studies is the ground state, an intermediate state is a core-excited state and a final state is a valence-excited state. The process that we consider involves absorption of two photons, an optical and an x-ray photon, an intermediate and a final state is a valence- and core-excited state, respectively. The resulting expressions for RIXS and our process involve different matrix elements. The similarity is that these matrix elements are expressed using the eigenvectors of the BSE wave function and momentum matrix elements between one-particle wave functions.
	
	\subsubsection{Independent particle approximation for optically-excited states}
	\label{Sec_IPA_optical}
	{ \eq{Eq_XAS_BSE} implies that both valence-excited states created by a pump pulse and core-excited states created by a probe pulse are obtained with the BSE. We now derive the signal by describing only core-excited states within the BSE and valence-excited states within the independent-particle approximation. The purpose of this derivation is to analyze, how a signal would change if excitonic effects are ignored for an optical excitation. Instead of an energy $E_{\lambda_I}$, we use the energy difference between one-particle energies of a state in a conduction and a valence band, $|e_{i}\kp\ra$ and $|h_{j}\kp\ra$, respectively. Each valence-excited state $\lambda_I$ is now characterized by only one excitation from a one-particle state $|h_{j}\kp\ra$ to a one-particle state $|e_{i}\kp\ra$. The coefficients $A^{\lambda_I}_{e h\kp}$ are then nonzero only for this excitation, $A^{\lambda_I}_{eh\kp} = \delta_{e_i,e}\delta_{h_j,h}$ and the cross section becomes
	\begin{widetext}
		\begin{align}
		\sigma = 
		\frac1c\sum_{\lambda_F} 
		\lf| \sum_{e,h,\kp} w_{eh} \frac{\sum_{\mu} A^{*\lambda_F}_{e \mu\kp} \int d^3 r \phi_{h\kp}^\dagger(\br)e^{i \mathbf k_x\cdot\br}( \boldsymbol\epsilon_x \cdot \hat {\mathbf p})  \phi_{\mu\kp}(\br)}{\Gamma_F - i((E_{e\kp}-E_{h\kp}) - E_{\lambda_F}+\omega_x)}
		\rt|^2,\label{Eq_IP}
		\end{align}
	\end{widetext}
with the weigths
\begin{align}
w_{eh} & = \sqrt{\frac{2\pi I_0}{c\omega_{\text{o}}^2}}\frac{ \sum_{e,h,\kp}\int d^3  \phi_{e}^\dagger(\br_2) (\boldsymbol\epsilon_{\text{o}}\cdot \hat{\mathbf p})\phi_{h} (\br_2)}{\Gamma_I - i(E_{0} - (E_{e{\kp}}-E_{h{\kp}})+\omega_{\text{o}})}.
\end{align}
Here, energies $E_{e_{i\kp}}$ and $E_{h_{j\kp}}$ are one-particle energies of a state in a conduction and valence band, respectively, which we obtain from a DFT calculation. We will use \eq{Eq_IP} to illustrate the role of an electron-hole coupling in valence-excited states for the signal calculated in Section \ref{SectionResults}.}

	\subsection{Computational details}
	
	We perform the BSE+G$_{0}$W$_{0}$ calculations for 4H-SiC implemented in \texttt{exciting} software package \cite{gulans2014exciting, draxl2017exciting} { version oxygen} to calculate both valence-excited states \cite{sagmeister2009time, puschnig2002optical, leng2016gw, vorwerk2019bethe, henneke2020fast} and core-excited states \cite{vorwerk2022all, urquiza2023pseudopotential, olovsson2009all, DEGROOT2021147061}. \texttt{exciting} is an electronic-structure density-functional-theory (DFT) package that employs the all-electron full-potential linearized augmented plane-wave method combined with local orbital approach (LAPW+lo) \cite{kutepov2021atomic}. This method allows for the treatment of core orbitals in solid-state materials \cite{unzog2022x}. A core state  $|\phi_{\mu\kp}\ra$ has a $\kp$-point index, since core states are represented as periodic Bloch functions within the LAPW+lo method. We use the generalized gradient approximation (GGA)\cite{stampfl2001electronic} in the implementation of Perdew, Burke, and Ernzerhof \cite{yuan2018gga} for exchange-correlation functional calculation in DFT calculations. G$_{0}$W$_{0}$ approximation was applied on top of the DFT calculation to correct quasi-particle energies that enter the BSE equation \cite{nabok2016accurate,salas2022electronic}. We performed calculations on a $4\times 4 \times 4$ $\mathbf k$-point grid, the number of unoccupied bands considered in G$_{0}$W$_{0}$ calculations were 300. Valence-excited and core-excited BSE calculations were performed using 18 unoccupied bands, with Lorentzian broadening widths of 0.48 eV and 0.32 eV, respectively. { The widths are adjusted to fit experimental optical and x-ray absorption spectra. See Appendix \ref{AppSpectra} for optical and x-ray spectra that can be compared with experimental data and Appendix \ref{App_ConvStudy} for the convergence study}. With the two BSE calculations, we obtain the necessary components for the evaluation of XAS spectra with \eq{Eq_XAS_BSE}: the coefficients $A^{\lambda_F}_{e \mu \kp}$ and $A^{\lambda_I}_{e v \kp}$, the energies $E_{\lambda_F}$ and $E_{\lambda_I}$, and the momentum matrix elements between the Kohn-Sham orbitals. { All input and output files are available at NOMAD \cite{NOMAD.dataset}}.

	\section{Results and discussion}

	\label{SectionResults}	
	We first analyze the optical absorption spectrum of 4H-SiC calculated with the BSE.  The absorption cross section averaged over polarizations is proportional to the imaginary part of the dielectric function $\varepsilon$ that is shown in Fig.~\ref{Fig_Ime}. 
	It agrees with the experimental data reported in Refs.~\cite{lindquist2001ordinary, harris1995properties}, (see appendix B for the data comparison). We also show $\operatorname{Im}(\varepsilon)$ calculated with the DFT and G$_{0}$W$_{0}$ for comparison in Fig.~\ref{Fig_Ime}.\\
	
	\begin{figure}[t]
		\centering
		\includegraphics[width=8.4cm, height=5.2cm]{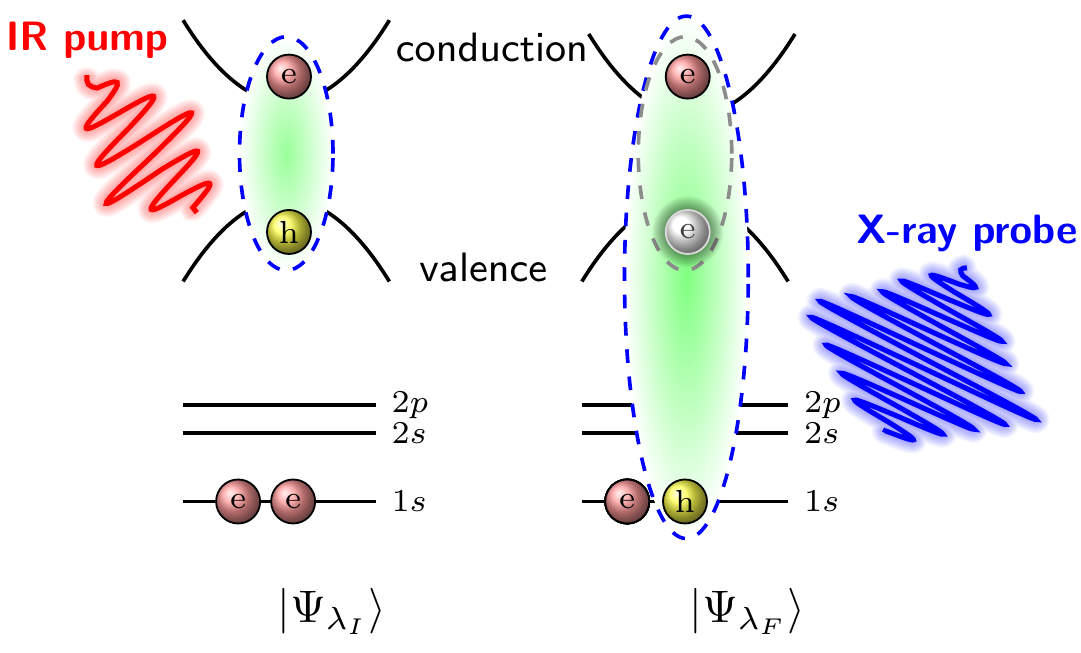}
		\caption{Scheme of the considered pump-probe experiment. First, a pump pulse creates a "valence exciton", where a red circle bounded to a yellow circle represents the created exciton. Second, an x-ray probe pulse creates a "core exciton" by inducing a transition of an electron from a core-shell to a vacancy in a valence band.}
		\label{Fig_Pump-Probe-Scheme}
	\end{figure}
	The significant difference between the DFT or G$_{0}$W$_{0}$ and the BSE+G$_{0}$W$_{0}$ optical spectra in the energy region 4--6 eV indicates that excitonic effects are particularly strong for excited states with these energies. For this reason, we further consider the photon energy of a pump pulse to be $4.6$ eV.	
	
	The properties of a pump pulse determine the shape of created excitons. In Fig.~\ref{Fig_BSE_optical_Ime}, we analyze how the structure of excitons depends on the pump-pulse polarization $\boldsymbol\epsilon_{o}$.  Figs.~\ref{Fig_BSE_optical_Ime}(b) - (d) show $|w_{\lambda_I}|^2$ for different valence-excited states $\lambda_I$ as a function of their energy $E_{\lambda_I}$. $|w_{\lambda_I}|^2$ is proportional to the probability of a transition to a $\lambda_I$ state. { The hole density of an excitonic state $\lambda_I$ is defined as $\rho^{\lambda_I}_h(\br_h) = \int d^3 r |\Psi_{\lambda_I}(\br_h,\br_e)|^2 d\br_e$ and can be obtained with the \texttt{exciting} software package. We weight each individual normalized hole distribution by the factor $|w_{\lambda_I}|^2$ and obtain the averaged hole density distribution of created excitons $\rho_h$ as
\begin{align}
\rho_h =\sum_{\lambda_I} |w_{\lambda_I}|^2 \rho^{\lambda_I}_h(\br_h),
\end{align} 
which is shown in Figs.~\ref{Fig_BSE_optical_Ime}(e) - (h) for different pump-pulse polarizations. Two-dimensional plane cuts of the averaged hole distributions are shown in Figs.~\ref{Fig_BSE_optical_Ime}(g) and (h). }

Comparing Figs.~\ref{Fig_BSE_optical_Ime}(b)-(d), we see that a pump pulse polarized along the $\mathbf a$ and a pump pulse polarized along the $\mathbf b$ lattice vector excite same excitonic states $\lambda_I$ with the same probability. We also obtain that the corresponding hole distributions shown in Fig.~\ref{Fig_BSE_optical_Ime}(e) and (g) are identical for these pump-pulse polarizations. They are predominantly of a Carbon $p$-type character aligned along the both $\mathbf a$ and $\mathbf b$ directions slightly bent towards the $\mathbf c$ vector. The exciton's hole distributions for $\boldsymbol\epsilon_{\text{o}}$ parallel to the $\mathbf c$ lattice vector are predominantly of a Carbon $p$-type character pointing along the $\mathbf c$ direction as shown in Fig.~\ref{Fig_BSE_optical_Ime}(f) and (h).
	
	\begin{figure}[tb]
	\centering
		\includegraphics[width=7.5cm, height=4.5cm]{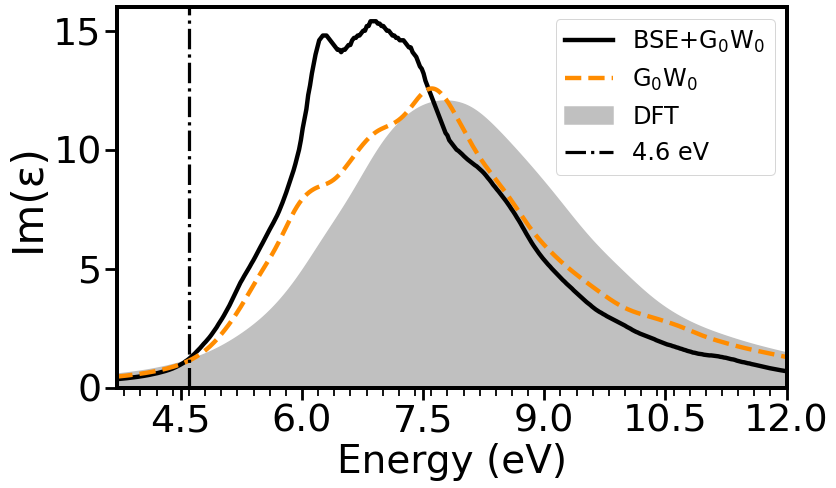}
		\caption{Imaginary part of the dielectric function of 4H-SiC.  Spectrum calculated by means of BSE+G$_{0}$W$_{0}$ is shown with a solid line.  Spectrum calculated by means of G$_{0}$W$_{0}$ is shown with a dashed orange line. Spectrum calculated by means of DFT is shown in the gray filled curve. The dotted line illustrates the photon energy of the pump pulse.}
		\label{Fig_Ime}
	\end{figure}
	
%XAS from 4H-SiC in the ground state in the shaded rectangles that is in a good agreement with experimental data \cite{tallarida2006x, luning1999electronic}
	
	We show the results of our calculations of Carbon K-edge and Si K-edge x-ray spectra of an optically-excited 4H-SiC probed by an x-ray pulse with an energy below XAT in Fig.~\ref{Fig_BSE_XAS_sigma}. We also show the calculated XAS from 4H-SiC in the ground state. As discussed in Section \ref{SecMethod}, the optical excitation leads to an appearance of an additional peak in the spectra in the pre-edge XAS region, where conventional XAS is vanishing. These emerged peaks are shown in red and blue for Carbon K-edge and Si K-edge in Fig.~\ref{Fig_BSE_XAS_sigma}, respectively. They appear due to the presence of a coupled electron-hole pair, in which the hole provides an additional channel for an electron transition from a core to a state near the Fermi energy. The position of the peak at an energy lower than XAT can be intuitively understood in a one-particle picture. In such a picture, the energy of the peak would be given by the energy difference between a core orbital and a valence band, which is lower than the energy of transitions to conduction bands involved in a ground-state XAS (see Fig.~\ref{Fig_Pump-Probe-Scheme}).
	\begin{widetext}
		\begin{minipage}{\linewidth}
			\begin{figure}[H]
				\includegraphics[width=3.8cm, height=7cm]{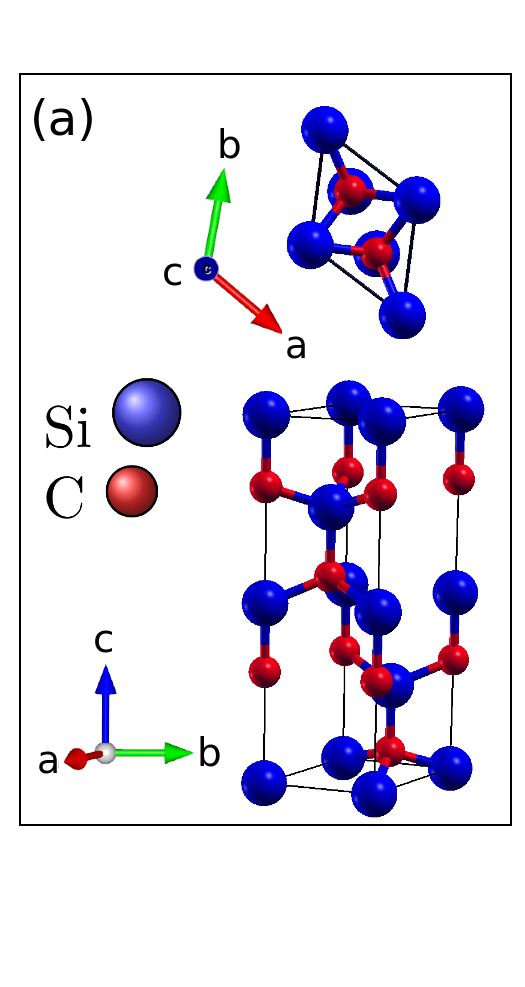}
				\includegraphics[width=13cm, height=7.8cm]{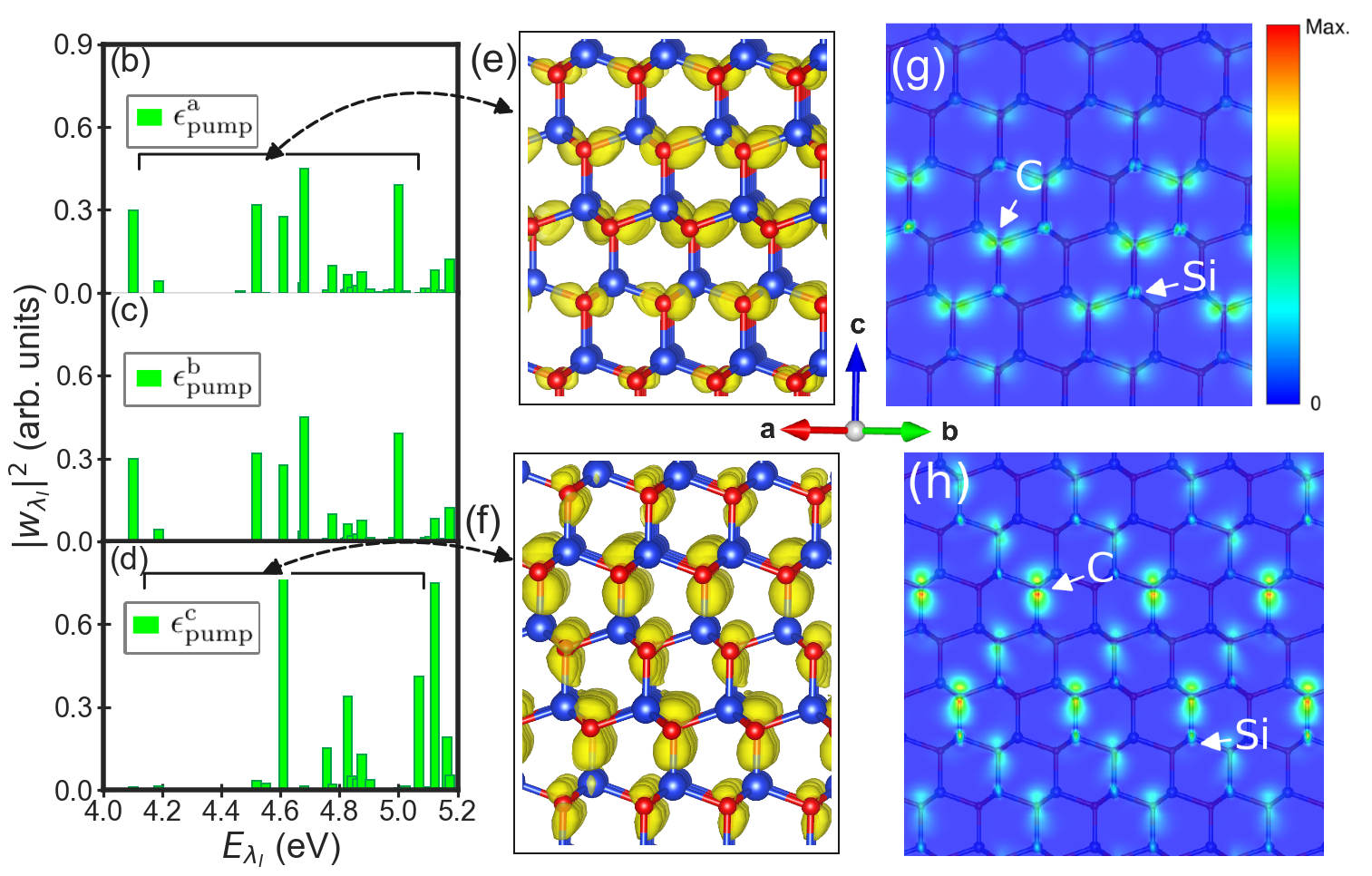}\hfill
				\centering 
				\caption{(a) Crystal structure of 4H-SiC. 
					(b)-(d) Weights $w_{\lambda_{I}}$ of valence-excited states $\lambda_I$ as a function of their energy $E_{\lambda_I}$ for the pump-pulse photon energy $\omega_{\text{o}}=4.6$ eV and polarization of a pump pulse along (b) the $\mathbf a$ lattice vector; (c) the $\mathbf b$ lattice vector and (d) the $\mathbf c$ lattice vector.
					(e)-(f) Averaged distribution of the excitons' holes in real space for a pump pulse polarized (e) perpendicular to the $\mathbf c$ lattice vector and (f) along $\mathbf c$ lattice vector. (g)-(h) The 2D-plane cuts of the averaged hole distribution of the excitonic states after the action of a pump pulse polarized (g) perpendicular to the $\mathbf c$ vector and (h) parallel to $\mathbf c$ vector. We used the VESTA package \cite{momma2008vesta} for visualization. }
				\label{Fig_BSE_optical_Ime}	
			\end{figure}	
		\end{minipage}
	\end{widetext}			
	
	%We applied our framework to calculate Carbon K-edge and Si K-edge spectra of a photo-excited 4H-SiC shown in Fig.~\ref{Fig_BSE_XAS_sigma}. We also show the XAS of the ground-state 4H-SiC, which are in a good agreement with the experiment (see SM) \cite{supplementary.material}, for comparison \cite{tallarida2006x, luning1999electronic}. 
	
	\begin{widetext}
		\begin{minipage}{\linewidth}
			\begin{figure}[H]
				\includegraphics[width=11cm, height=8.9cm]{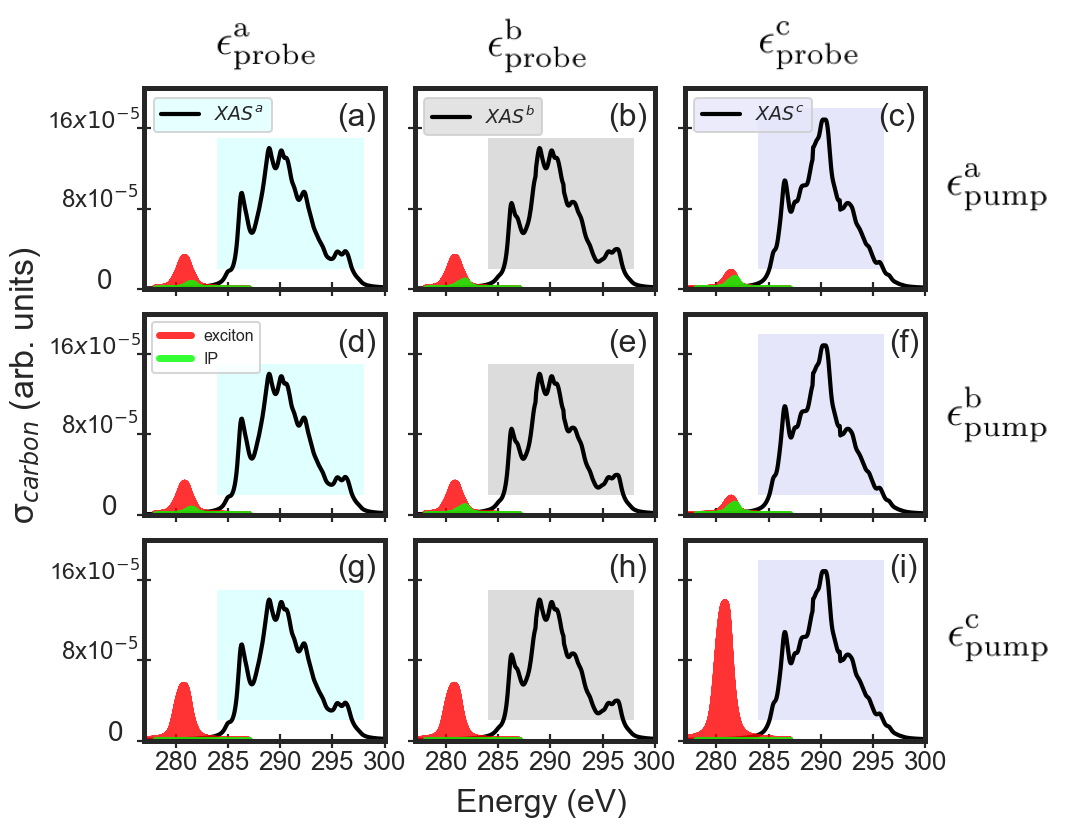}
				\includegraphics[width=4.1cm, height=8.7cm]{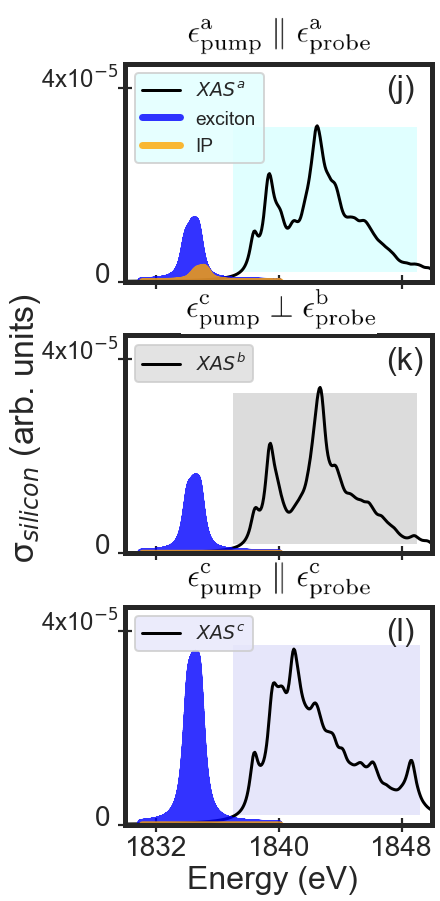}
				%\centering
				\caption{Carbon and Silicon K-edges XAS spectra. Red and green peaks show XAS spectra of optically-excited 4H-SiC at Carbon and Silicon Kedges, respectively, for different polarizations of a pump and a probe pulse. { Green and orange peaks are the same as the red and green peaks, respectively, but calculated using the independent particle approximation for the optical excitation.} Black curves shaded in rectangles show XAS spectra of a ground-state 4H-SiC depending on the probe-pulse polarization. { All ground-state spectra are scaled to the same arbitrary unit. XAS spectra of optically-excited 4H-SiC are scaled by the factor $1\text{ a.u.}/(20\cdot I_0)$ relative to the ground-state spectra.} (a)--(i) Carbon K-edge XAS spectra for (a) $\boldsymbol\epsilon_{\text{o}}\parallel\mathbf a$, $\boldsymbol\epsilon_x\parallel\mathbf a$; (b) $\boldsymbol\epsilon_{\text{o}}\parallel\mathbf a$, $\boldsymbol\epsilon_x\parallel\mathbf b$;   {(c)} $\boldsymbol\epsilon_{\text{o}}\parallel\mathbf a$, $\boldsymbol\epsilon_x\parallel\mathbf c$;
					{(d)} $\boldsymbol\epsilon_{\text{o}}\parallel\mathbf b$, $\boldsymbol\epsilon_x\parallel\mathbf a$; {(e)} $\boldsymbol\epsilon_{\text{o}}\parallel\mathbf b$, $\boldsymbol\epsilon_x\parallel\mathbf b$;  {(f)} $\boldsymbol\epsilon_{\text{o}}\parallel\mathbf b$; $\boldsymbol\epsilon_x\parallel\mathbf c$;  {(g)} $\boldsymbol\epsilon_{\text{o}}\parallel\mathbf c$; $\boldsymbol\epsilon_x\parallel\mathbf a$; (h) $\boldsymbol\epsilon_{\text{o}}\parallel\mathbf c$, $\boldsymbol\epsilon_x\parallel\mathbf b$; (i) $\boldsymbol\epsilon_{\text{o}}\parallel\mathbf c$, $\boldsymbol\epsilon_x\parallel\mathbf c$. (j)--(l) Silicon K-edge XAS spectra for the polarization of pump and probe being parallel to (j) $\boldsymbol\epsilon_{\text{o}}\parallel\mathbf a$; $\boldsymbol\epsilon_x\parallel\mathbf a$; (k) $\boldsymbol\epsilon_{\text{o}}\parallel\mathbf c$, $\boldsymbol\epsilon_x\parallel\mathbf b$; (l) $\boldsymbol\epsilon_{\text{o}}\parallel\mathbf c$, $\boldsymbol\epsilon_x\parallel\mathbf c$.  }
				\label{Fig_BSE_XAS_sigma}	
			\end{figure}
		\end{minipage}
	\end{widetext}

	{ Intensities of the emerged peaks relative to the intensity of the ground-state signal are linearly proportional to the intensity of an optical pulse $I_{\text{o}}$. We magnified the intensities in Fig.~\ref{Fig_BSE_XAS_sigma} for illustration purposes. For a correct relative scale at a given intensity $I_{\text{o}}$, the peaks intensities must be multiplied by a factor of $\sim 20 \cdot I_{\text{o}}/\text{a.u.}$, where $I_{\text{o}}$ is taken in atomic units. }

	\begin{figure}[tb]
	\centering
		\includegraphics[width=0.48\textwidth]{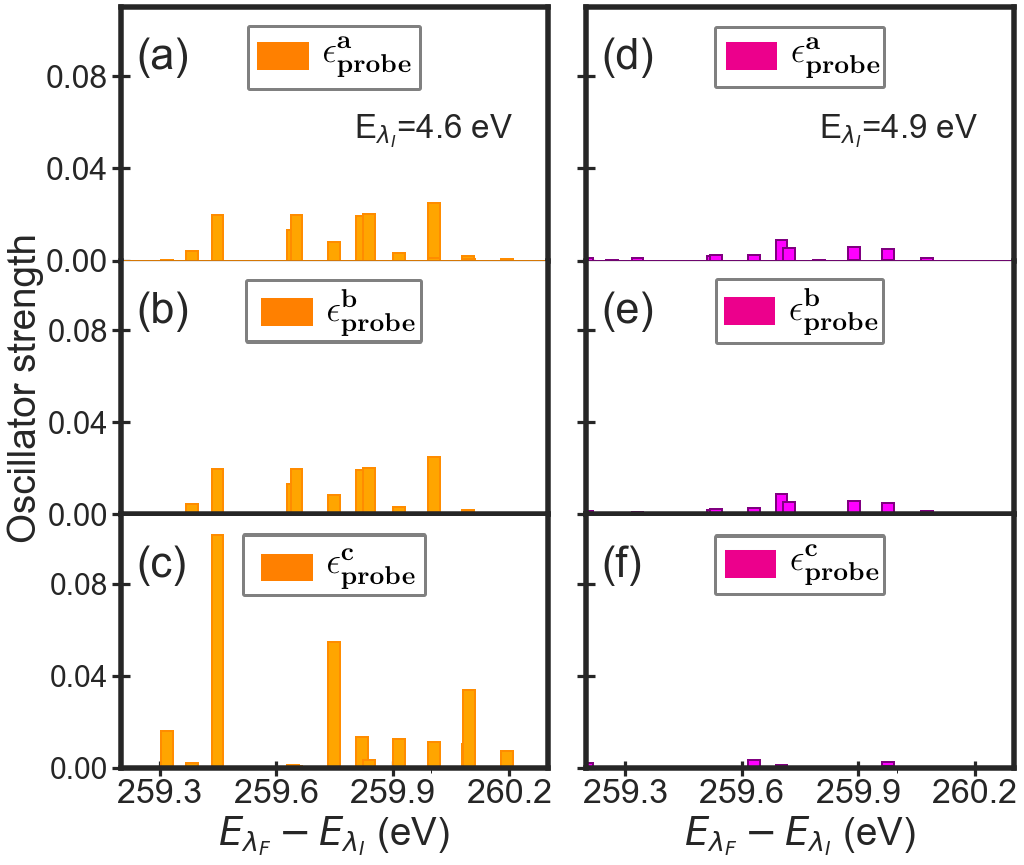}
		\caption{{ Oscillator strengths of transitions from a state $\lambda_I$ to states $\lambda_F$ for different probe-pulse polarizations. $\lambda_I$ with an energy 4.6 eV and probe-pulse polarizations parallel to (a) $\mathbf a$; (b) $\mathbf b$ and (c) $\mathbf c$ crystal axes are selected for the left column. $\lambda_I$ with an energy 4.9 eV and probe-pulse polarizations parallel to (d) $\mathbf a$; (e) $\mathbf b$ and (f) $\mathbf c$ crystal axes are selected for the right column.}}
		\label{Fig_States_F}
	\end{figure}
	
	Intensities of the peaks depend strongly on the polarization of a pump and a probe pulse. The polarization of a pump pulse determines a shape of a valence-excited exciton, which in turn determines selection rules for transitions induced by a probe pulse. XAS spectra of 4H-SiC excited by a pump pulse with polarizations $\boldsymbol\epsilon_{\text{o}}\parallel \mathbf a$ are shown in Figs.~\ref{Fig_BSE_XAS_sigma}(a)-(c). K-edge XAS spectra are sensitive to a $p$-type contribution to a wave function of unoccupied states, since they are governed by transitions from a $1s$ orbital. In agreement with our observation that the hole is of a predominantly $p$-type character pointing symmetrically in the $\mathbf a$ and $\mathbf b$ directions, we observe that intensive and identical peaks appear for the probe pulses polarized along the $\mathbf a$ and along the $\mathbf b$ direction. A small peak appears in the case of $\boldsymbol\epsilon_x\parallel \mathbf c$ as well, which is because the hole has also a non-zero projection along the $\mathbf c$ direction as can be seen in the two-dimensional (2D) cuts of the density in the Figs.~\ref{Fig_BSE_optical_Ime}(h) and (i). The intensities of the peaks for a pump pulse polarized along the $\mathbf b$ direction in Fig.~\ref{Fig_BSE_XAS_sigma}(d)-(f) are the same as that in Fig.~\ref{Fig_BSE_XAS_sigma}(a)-(c) in agreement with the fact that the averaged hole distributions are identical for the pump pulses polarized along $\mathbf a$ and along $\mathbf b$. 
	
	XAS signals for 4H-SiC excited by an optical pulse polarized along the $\mathbf c$ direction are shown in Fig.~\ref{Fig_BSE_XAS_sigma}(g)-(i). The strongest XAS signal in this case is for the probe pulse polarized along the $\mathbf c$ direction. This indicates that the hole distribution is predominantly aligned along the $\mathbf c$ vector, which agrees with our observation in Fig.~\ref{Fig_BSE_optical_Ime}(f). These results demonstrate that there is a direct correspondence between the shape of the excitons' hole distribution and the intensity dependence of the corresponding feature in the XAS spectra on the polarization of the x-ray probe pulse.  
	
	The peaks also emerge in the pre-edge region of Si K-edge XAS spectra as shown in Fig.~\ref{Fig_BSE_XAS_sigma}(j)-(l). The intensities of these peaks are overall smaller than the intensities of the peaks in pre-edge region of Carbon K-edge XAS spectra. This indicates that the hole has a contribution of a $p$-type character centred on Si atoms, which is smaller in comparison to the contribution centered on C atoms. The 2D density cuts in Figs.~\ref{Fig_BSE_optical_Ime}(h) and (i) confirm this conclusion.

The central positions of the peaks in the pre-edge region vary slightly depending on polarizations of a pump and a probe pulses as shown in Table \ref{TablePositions}. Variation in the central positions are due to the fact that each peak in the pre-edge region consists of a series of peaks. These peaks are centred at energies $E_{\lambda_F}-E_{\lambda_I}-\omega_x$ and their positions do not change with polarizations of the pump and the probe pulse. But the intensity of each peak does vary, because its is determined by a product of a transition amplitude describing an excitation from the ground state to an intermediate state $\lambda_I$ with a transition amplitude describing an excitation from the state $\lambda_I$ to a state $\lambda_F$. Fig.~\ref{Fig_BSE_optical_Ime} (b)-(d) shows that there is a number of states $\lambda_I$ with various energies $E_{\lambda_I}$ with a considerable corresponding transition amplitude. These transition amplitudes depend on the polarization of the pump pulse. In addition, the signal depends on the transition amplitudes that describes  excitations from states $\lambda_I$ into final states $\lambda_F$. Fig.~\ref{Fig_States_F} shows oscillator strengths for transitions from two selected intermediate states $\lambda_I$ to different states $\lambda_F$. These two states were selected as an example to demonstrate that the amplitudes of transitions to a state $\lambda_F$ depend on an intermediate state and polarization of the probe pulse. This way the broad peak in the pre-edge region is formed by a number of individual peaks with intensities that depend on polarizations of the pump and the probe pulse. Variation in intensities of individual peaks leads to the shift of the center of the broad peak.

%\begingroup
%\squeezetable
\begin{table}
\caption{\label{tab:ex}{ Central positions of the peaks in the pre-edge region of Carbon and Silicon K-edge XAS shown in Fig.~\ref{Fig_BSE_XAS_sigma} relative to XAT depending on polarizations of a pump and a probe pulse.}}
\begin{ruledtabular}
Carbon K-edge\\
\begin{tabular}{llll}
  & $\boldsymbol\epsilon_x\parallel\mathbf a$ & $\boldsymbol\epsilon_x\parallel\mathbf b$ &$\boldsymbol\epsilon_x\parallel\mathbf c$  \\
$\boldsymbol\epsilon_{\text{o}}\parallel\mathbf a$ & (a)  4.0495 eV & (b) 4.0495 eV & (c) 3.8295 eV  \\
$\boldsymbol\epsilon_{\text{o}}\parallel\mathbf b$ & (d) 4.0495 eV & (e) 4.0495 eV & (f) 3.8295 eV \\
$\boldsymbol\epsilon_{\text{o}}\parallel\mathbf c$ & (g) 4.1395 eV& (h) 4.1395 eV & (i) 4.1395 eV  \\
\end{tabular}
Silicon K-edge\\
\begin{tabular}{ll}
$\boldsymbol\epsilon_{\text{o}}\parallel\mathbf a$, $\boldsymbol\epsilon_x\parallel\mathbf a$ &  (j) 3.8228 eV  \\
$\boldsymbol\epsilon_{\text{o}}\parallel\mathbf c$, $\boldsymbol\epsilon_x\parallel\mathbf b$ &  (k) 3.8228 eV\\
$\boldsymbol\epsilon_{\text{o}}\parallel\mathbf c$, $\boldsymbol\epsilon_x\parallel\mathbf c$ &  (l) 3.8534 eV
 \end{tabular}
\end{ruledtabular}
\label{TablePositions}
\end{table}

{ To highlight the role of many-body effects for the XAS spectra of an optically-excited material, we additionally calculate the signal under the independent-particle approximation for a valence excitation derived in Sec.~\ref{Sec_IPA_optical}. Namely, we use \eq{Eq_IP} and obtain valence-excited states with the DFT instead of the BSE and core-excited states are obtained with the BSE. Peaks calculated with this assumption are shown in green for C K-edge in Figs.~\ref{Fig_BSE_XAS_sigma}(a)-(i) and in orange for Si K-edge in Figs.~\ref{Fig_BSE_XAS_sigma}(j)-(l). We obtain that the intensity of the peaks is lower in this case. We also find that the dependence of their intensity on polarizations of the pump and probe pulses disagrees with our former calculation. In the independent-particle approximation, there is no signal for a pump pulse aligned along the $\mathbf c$ direction. This illustrates that the independent particle approximation can provide qualitative errors in the description of the process. If many-body effects play an important role for optically-excited states, it is necessary to take them into account for a correct description of XAS spectra from a photo-excited material.}

	\section{Conclusions}
	\label{SectionConclusions}
	
	In conclusion, we developed an ab initio framework to describe optical-pump--x-ray-probe experiments in solid-state materials that treats excitonic effects at the BSE level, the state-of-the-art many-body approach. This way, it is capable to accurately describe modern experiments, where excited-state dynamics in functional materials or complexes are probed by means of x-ray pulses \cite{wernet2019chemical,GarrattStructuralDynamics24}. Using the example of 4H-SiC, we demonstrated that a shape of an exciton can strongly depend on the excitation properties, which can eventually affect exciton transport. XAS is a powerful tool to reveal details about the shape of a positive charge of a photo-excited material with an element- and orbital-specificity. 
	
	%{\green I like this sentence { This observation underscores the importance of considering crystallographic symmetry and polarization dependencies when studying excitonic properties in materials with specific symmetries.}, but it does not really match to the text at the moment. Maybe, we incorporate it later.}
	
	\section*{Acknowledgement}
	We gratefully acknowledge the financial support of the Volkswagen Foundation grant number 96237.
	
	\makeatletter
	\appto{\appendix}{%
		\@ifstar{\def\thesection{\unskip}\def\theequation@prefix{A.}}%
		{\def\thesection{\Alph {section}}}%
	}
	\makeatother
	\appendix
	
	\section{Derivation of an X-ray absorption cross section within the second-order time-dependent perturbation theory}	
	We evaluate the absorption cross section in the considered pump-probe experiment using the second-order time-dependent perturbation theory. The interaction Hamiltonian between the electron system and light is
	\begin{align}
		\Hint (t) = \frac1c \int d^3 r \hat\psi^\dagger (\br)(\mathbf A(\br,t)\cdot \hat{\mathbf p}) \psi(\br),
	\end{align}
	where $c$ is the speed of light, $\hat\psi$ is the field annihilation operator, $\hat{\mathbf p}$ is the momentum operator and the $\mathbf A(\br,t)$ is the vector potential given by
	\begin{align}
		\mathbf A(\br,t) = c \frac {\mathcal {\mathbf E}_o(\br,t)}{\omega_{\text{o}}} \cos(\omega_{\text{o}} t ) + c \frac {\mathcal {\mathbf E}_x(\br,t)}{\omega_x} \cos(\omega_x t),
	\end{align}
	Here, $\omega_x$ and $\omega_{\text{o}}$ are the photon energies of the x-ray and optical pulse, respectively. $\mathcal {\mathbf E}_o(\br,t)$ and $\mathcal {\mathbf E}_x(\br,t)$ are the electric fields of the x-ray and optical pulse, respectively. The second-order wave function after the interaction with light is
\begin{widetext}
		\begin{align}
			|\Psi_2\ra = &\sum_{\lambda_I,\lambda_F}\int_{-\infty}^{\infty} d t_1 \int_{-\infty}^{t_1} d t_2 e^{-i(E_{\lambda_I} - E_{\lambda_F})t_2 }e^{-\Gamma_F|t_2|}e^{-i(E_{0} - E_{\lambda_I})t_1 }e^{-\Gamma_I|t_1|} \\
			&\times\la \Psi_{\lambda_F}| \Hint(t_1)| \Psi_{\lambda_I}\ra \la  \Psi_{\lambda_I}|  \Hint(t_2)| \Psi_0\ra \nonumber.
		\end{align}
	The initial state $| \Psi_0\ra$ is the ground state, the states after the action of the optical pulse and after the action of the x-ray pulse are states $| \Psi_{\lambda_I}\ra$ and $| \Psi_{\lambda_F}\ra$, respectively. They are valence- and core-excited state described by the wave function defined below
	\begin{align}
		|\Psi_\lambda \ra = \sum_{e,h,\kp} A^\lambda_{e h\kp}| e_\kp h_\kp \ra.\label{eq_BSE_wf}%=  \sum_{e_\kp,h_\kp} A^\lambda_{vc}c_{e_\kp}^\dagger \hat c_{h_\kp} | \Psi_0\ra,
	\end{align}
	We introduced the factors $\Gamma_I$ and $\Gamma_F$ that account for the finite life time of the excited states. We continue further with the derivation and obtain
		\begin{align}
			|\Psi_2\ra =& \frac {\mathcal E_o\mathcal  E_x }{4\omega_x\omega_0}   \sum_{\lambda_I,\lambda_F} \int_{t'''}^{t''''}  d t_2  e^{-i(E_{\lambda_I} - E_{\lambda_F}+\omega_x)t_2 } e^{-\Gamma_F|t_2|}\int_{t'}^{t''} d t_1 e^{-i(E_{0} - E_{\lambda_I}+\omega_{\text{o}})t_1 } e^{-\Gamma_I|t_1|} \\
			&\times\la \Psi_{\lambda_F}  |\hat \psi^\dagger(\br_1) e^{i\mathbf k_x\cdot \br_1}(\boldsymbol\epsilon_x\cdot \hat{\mathbf p})\hat \psi (\br_1) |\Psi_{\lambda_I}\ra \la  \Psi_{\lambda_I}|  \hat \psi^\dagger(\br_2) (\boldsymbol\epsilon_{\text{o}}\cdot \hat{\mathbf p})\hat \psi (\br_2) | \Psi_0\ra \nonumber,
		\end{align}
	where $\mathcal E_o$ and $\mathcal  E_x$ are the amplitudes of the corresponding electric fields, and $\epsilon_0$ and $\epsilon_x$ are polarization directions of the optical and x-ray pulse, respectively.  The pulses are assumed to not overlap in time.  We employ the dipole approximation for the description of the interaction with the optical pulse. We now make use of the assumption that the pulses have a long duration such that their  time dependence can be ignored for the evaluation of the integrals
		\begin{align}
			|\Psi_2\ra =& \frac {\mathcal E_o\mathcal  E_x }{4\omega_x\omega_0}   \sum_{\lambda_I,\lambda_F}\frac{ \la \Psi_{\lambda_F}  |\hat \psi^\dagger(\br_1) e^{i\mathbf k_x\cdot \br_1}(\boldsymbol\epsilon_x\cdot \hat{\mathbf p})\hat \psi (\br_1) |\Psi_{\lambda_I}\ra }{\Gamma_F - i(E_{\lambda_I} - E_{\lambda_F}+\omega_x)} 
			\frac{ \la  \Psi_{\lambda_I}|  \hat \psi^\dagger(\br_2) (\boldsymbol\epsilon_{\text{o}}\cdot \hat{\mathbf p})\hat \psi (\br_2) | \Psi_0\ra}{\Gamma_I - i(E_{0} - E_{\lambda_I}+\omega_{\text{o}})}. \label{Eq_Supp_2ndwf}
		\end{align}
	We now evaluate the matrix element $\la \Psi_{\lambda_F}  |\hat \psi^\dagger(\br) e^{i\mathbf k_x\cdot \br_1}(\boldsymbol\epsilon_x\cdot \hat{\mathbf p})\hat \psi (\br) |\Psi_{\lambda_I}\ra$ using the definition of the field operators in the main text
		\begin{align}
			\la \Psi_{\lambda_F}  |\hat \psi^\dagger(\br_1) e^{i\mathbf k_x\cdot \br_1}(\boldsymbol\epsilon_x\cdot \hat{\mathbf p})\hat \psi (\br_1) |\Psi_{\lambda_I}\ra
			= &\sum_{e_1,h_1,\kp_1, e_2,h_2,\kp_2,\mu,v,\kp,\kp' } A^{*\lambda_F}_{e_2 h_2\kp_2} A^{\lambda_I}_{e_1 h_1\kp_1} \int d^3 r \phi_{v\kp'}^\dagger(\br)e^{i \mathbf k_x\cdot\br}( \boldsymbol\epsilon \cdot \hat {\mathbf p})  \phi_{\mu\kp}(\br)  \label{Eq_Supp_matrixelement1}
			\\
			& \times\la \Psi_0^{\Nel}|  c_{e_{2}\kp_2}^\dagger c_{h_{2}\kp_2}  \hat c_{v\kp'}^\dagger  \hat c_{\mu\kp} c_{e_{1}\kp_1}^\dagger \hat c_{h_{1}\kp_1}  | \Psi_0^{\Nel}\ra  \nonumber.
		\end{align}
	Let us evaluate the product of the matrix element and the integral $t^x_{v\mu}=\int d^3 r \phi_{v\kp'}^\dagger(\br)e^{i \mathbf k_x\cdot\br}( \boldsymbol\epsilon \cdot \hat {\mathbf p})  \phi_{\mu\kp}(\br) $ in \eq{Eq_Supp_matrixelement1}. By assumption, the final state is a core-excited state and that's why only the basis states, where the state $|\phi_{h_{2}\kp_2}\ra$ is a core state, enter into the two-particle wave function. The initial state is a valence-excited state and the state $|\phi_{h_{1}\kp_1}\ra$ is a state in a valence band. That is why, the states   $|\phi_{h_{2}\kp_2}\ra$ and  $|\phi_{h_{1}\kp_1}\ra$ cannot coincide by assumption. If the states $|\phi_{p\kp}\ra$ and $|\phi_{v\kp'}\ra$ coincide, then the integral $t^x_{v\mu}$ would be zero. Thus, the product  of the matrix element and the integral is simply proportional to the product of six Kronecker delta functions
		\begin{align}
			&\la \Psi_0^{\Nel}|  c_{e_{2\kp_2}}^\dagger c_{h_{2\kp_2}}  \hat c_{v_\kp}^\dagger  \hat c_{\mu_\kp} c_{e_{1\kp_1}}^\dagger \hat c_{h_{1\kp_1}}  | \Psi_0^{\Nel}\ra \int d^3 r \phi_{v_\kp}^\dagger(\br)e^{i \mathbf k_x\cdot\br}( \boldsymbol\epsilon \cdot \hat {\mathbf p})  \phi_{\mu_\kp}(\br)  \\
			&=\delta_{\mu, h_2} \delta_{\kp, \kp_2} \delta_{e_{2}, e_{1}} \delta_{\kp_2, \kp_1} \delta_{v, h_1}  \delta_{\kp, \kp_1}  \int d^3 r \phi_{v\kp'}^\dagger(\br)e^{i \mathbf k_x\cdot\br}( \boldsymbol\epsilon \cdot \hat {\mathbf p})  \phi_{\mu\kp}(\br) \nonumber .
		\end{align}
We substitute this in \eq{Eq_Supp_matrixelement1} and obtain
\begin{align}
\la \Psi_{\lambda_F}  |\hat \psi^\dagger(\br_1) e^{i\mathbf k_x\cdot \br_1}(\boldsymbol\epsilon_x\cdot \hat{\mathbf p})\hat \psi (\br_1) |\Psi_{\lambda_I}\ra = \sum_{e,v,\mu,\kp } A^{*\lambda_F}_{e \mu\kp} A^{\lambda_I}_{e v\kp} \int d^3 r \phi_{v\kp}^\dagger(\br)e^{i \mathbf k_x\cdot\br}( \boldsymbol\epsilon_x \cdot \hat {\mathbf p})  \phi_{\mu\kp}(\br) .
\end{align}

\end{widetext}
\newpage

	\section{Optical and x-ray absorption spectra of 4H-SiC}
	\label{AppSpectra}
	
	Fig.~\ref{optical_z} (a)-(b) shows the imaginary part of the dielectric function for an optical pump polarized parallel to the $c$ axis and perpendicular to the $c$ axis. We find it to be in a good agreement with the experimental data reported in \cite{lindquist2001ordinary}.
	\begin{figure}[tb]
		\centering
		\minipage{0.42\textwidth}
		\includegraphics[width=.77\linewidth]{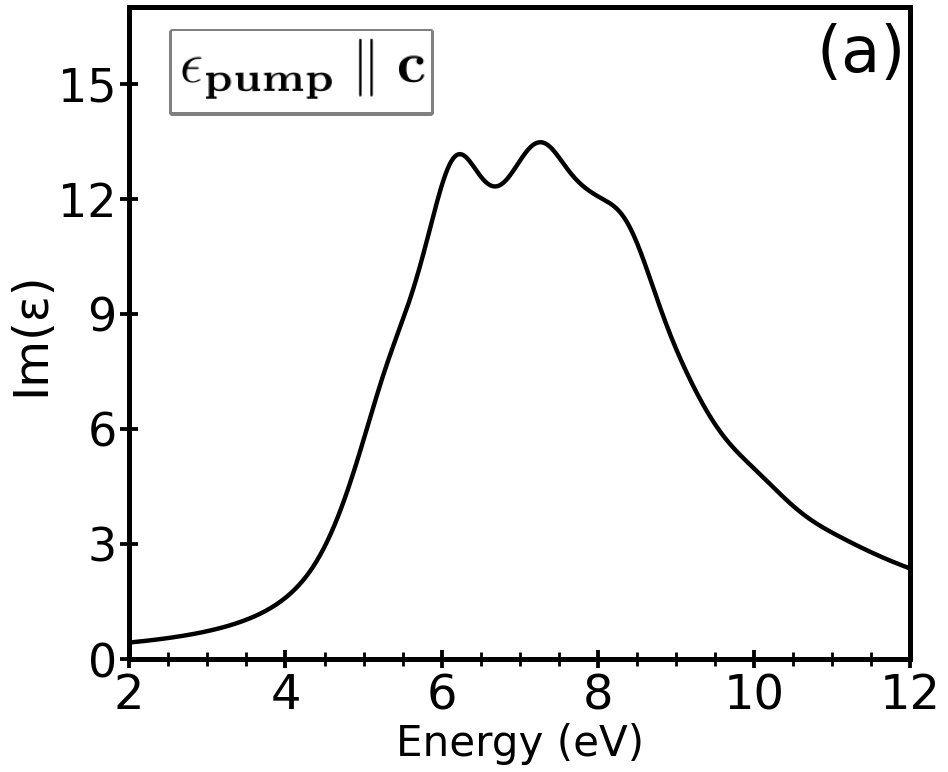}
		\endminipage \hfill\hfill
		\minipage{0.42\textwidth}
		\includegraphics[width=.77\linewidth]{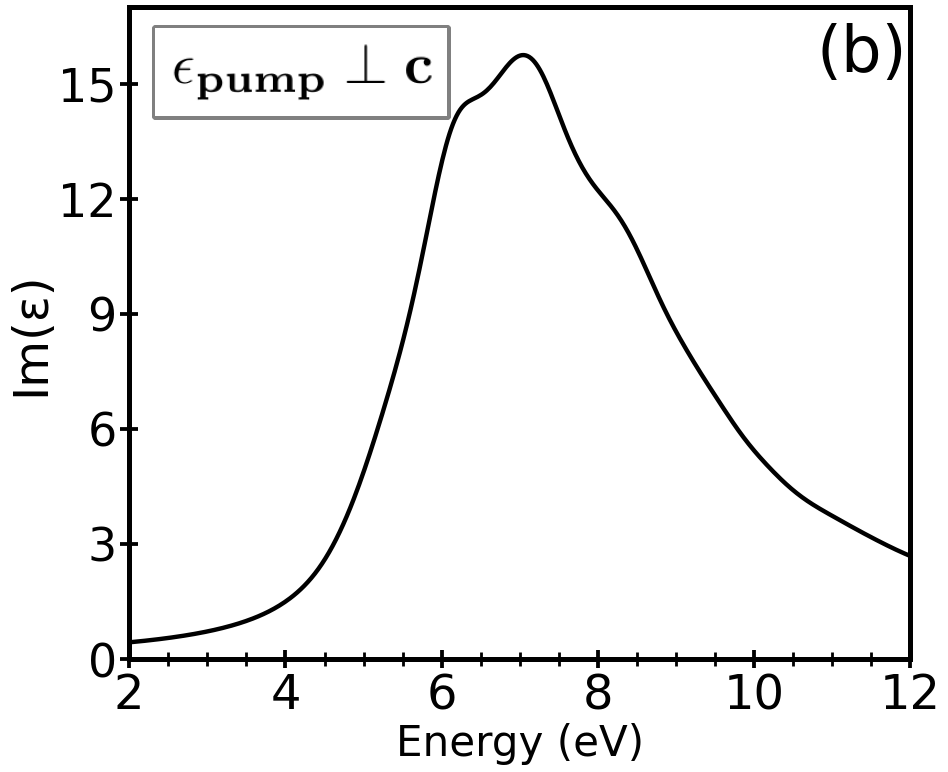}
		\endminipage
		\caption{Imaginary part of the dielectric function of 4H-SiC for the pump pulse polarized  (a)  parallel  to $\mathbf c$ and  (b)  perpendicular to $\mathbf c$.}
		\label{optical_z}
	\end{figure}	
	
	Figs.~\ref{C_Si_Kedges} (a)-(b) show polarization-averaged XAS spectra of 4H-SiC in the ground state. Carbon K-edge spectra are in a good agreement with the experimental data reported in Refs.~\cite{tallarida2006x, luning1999electronic} in the energy range 284 -- 296 eV. The theoretical and experimental spectra cannot be compared at higher energies, since the energy range of spectra is limited by the number of considered empty states in the theory. We are not aware of experimental data for Silicon K-edge, which could be used for the comparison to our theoretical results.
	
	\begin{figure}[tb]
		\centering
		\minipage{0.40\textwidth}
		\includegraphics[width=.81\linewidth]{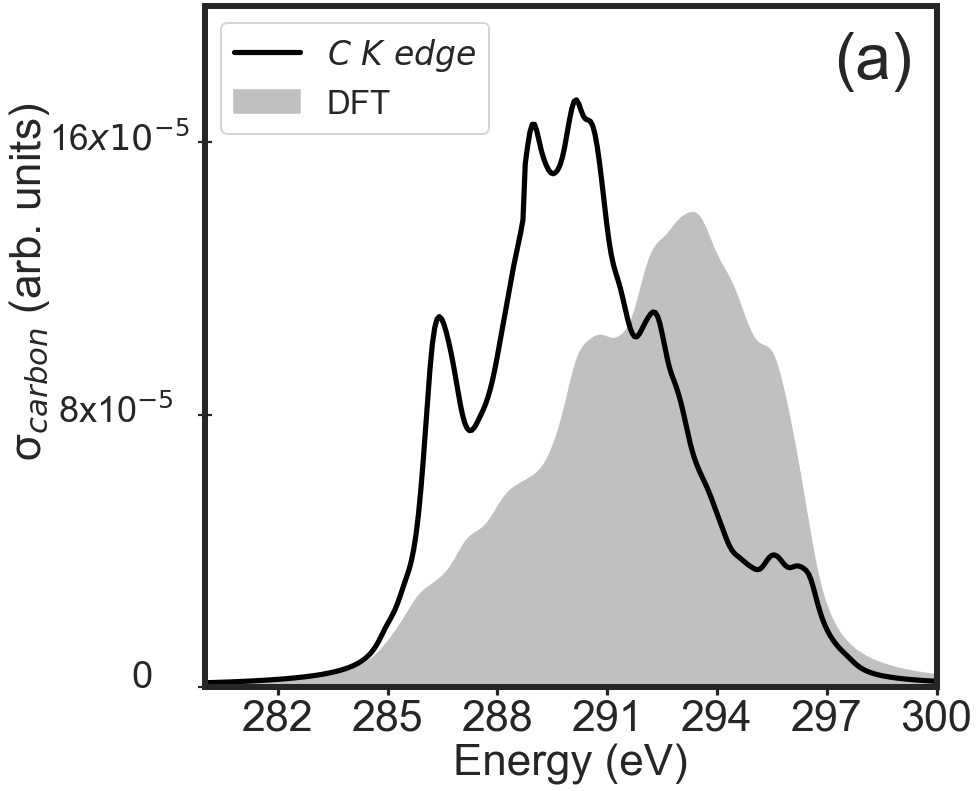}
		\endminipage \hfill\hfill
		\minipage{0.40\textwidth}
		\includegraphics[width=.84\linewidth]{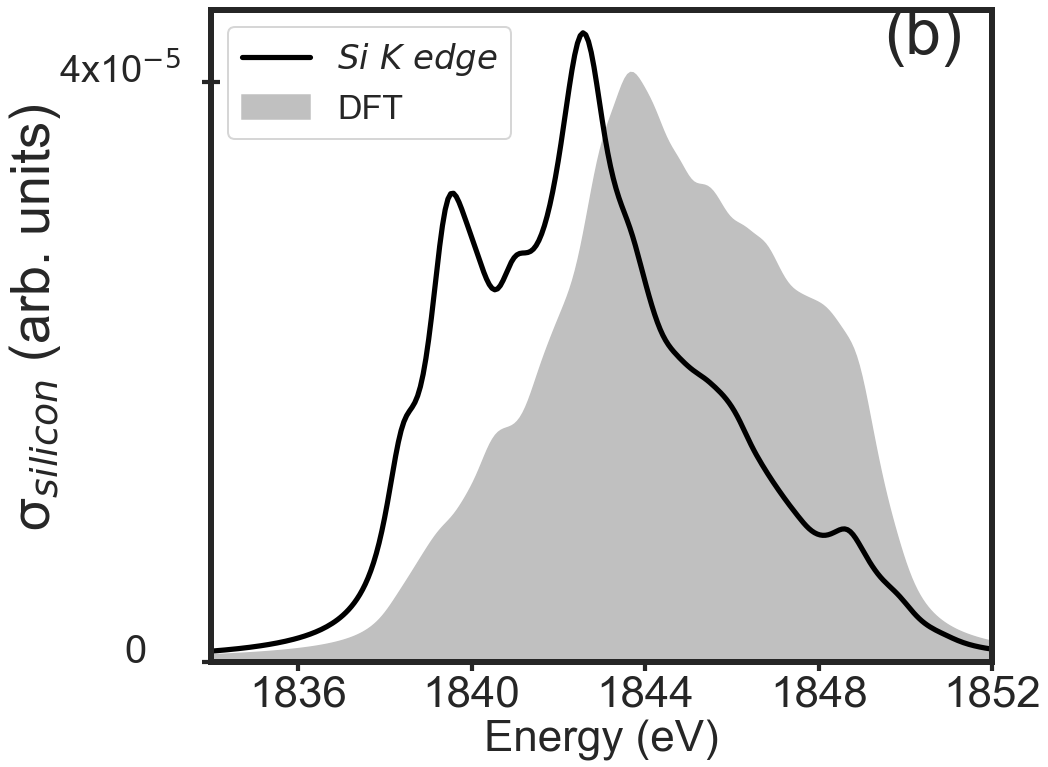}
		\endminipage
		\caption{The XAS absorption cross section of 4H-SiC in the ground state averaged over polarizations calculated by means of BSE+G$_{0}$W$_{0}$ { is shown with a solid line and by means of DFT is shown in the gray filled curve} for (a) Carbon K edge and  (b) Si K-edge.}
		\label{C_Si_Kedges}
	\end{figure}

\section{ Convergence study} 
\label{App_ConvStudy}

{ Fig.~\ref{GW.convergence} shows the band gap calculated with G$_{0}$W$_{0}$ depending on the size of the $\kp$-point grid and on the number of empty bands. We first fix the number of empty bands to be 300 and converge the size of the $\kp$-point grid. We then fix $4\times 4 \times 4$ for the $\kp$-point grid and check the convergence behaviour with respect to the number of empty bands. The $\kp$-grid size used in the calculations with the G$_{0}$W$_{0}$ and with the BSE must coincide.  In Fig.~\ref{BSE.convergence}, we study how optical absorption spectra calculated with the BSE depend on the $\kp$-grid size. Due to a high computational cost of calculations, we need to limit our selection to the smallest grid size, which provides reasonable results, which is $4\times 4 \times 4$ grid.}

\begin{figure}[tb]
	\centering
	\includegraphics[width=7.5cm, height=5cm]{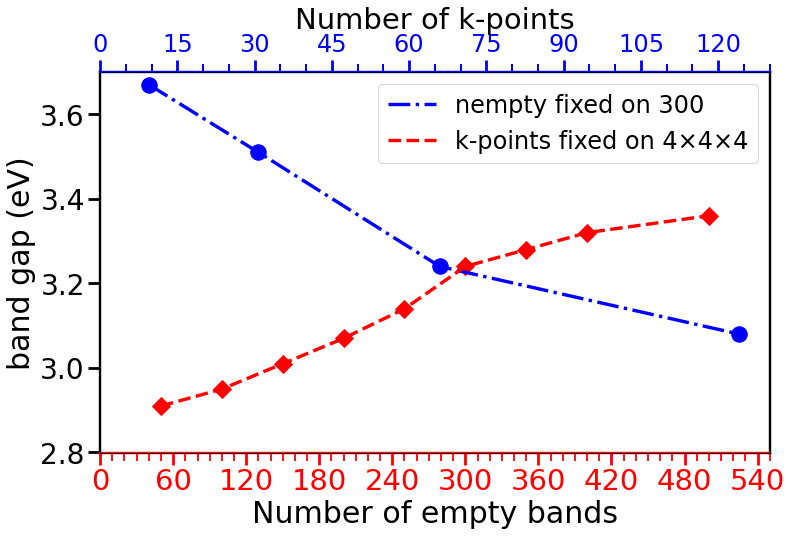}
	\caption{Dependence of the band gap calculated with G$_{0}$W$_{0}$ on the size of the $\kp$-point grid in blue and on the number of empty bands in red.}
	\label{GW.convergence}
\end{figure}

%{\green Additionally, we have showed the exciton binding energy as a function of the number of $\mathbf k$-points, marking in each case the corresponds grid size. }
\begin{figure}[tb]
		\centering
		\includegraphics[width=8.1cm, height=5.1cm]{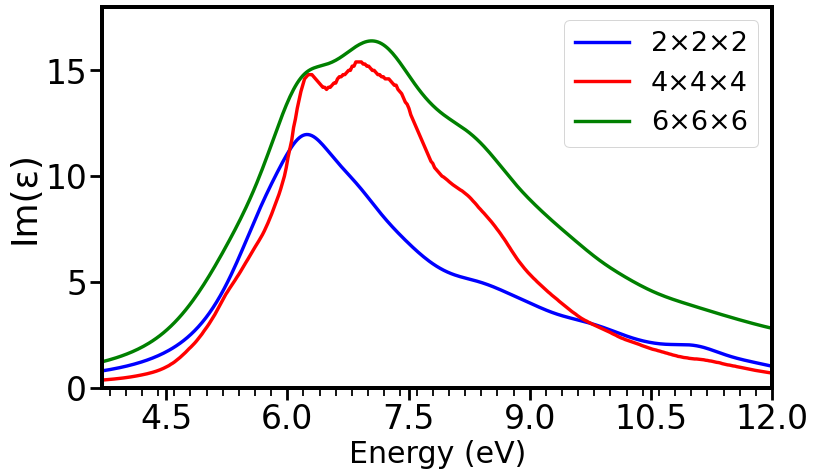}
		\caption{Imaginary part of the dielectric function calculated with the BSE at different $\mathbf k$-point grids.}
		\label{BSE.convergence}
	\end{figure}
 \clearpage


%	\bibliography{refs.bib}

\end{document}